\documentclass[aps,prb,twocolumn,superscriptaddress]{revtex4} 
\usepackage{graphicx}  % needed for figures
\usepackage{dcolumn}   % needed for some tables
\usepackage{bm}        % for math
\usepackage{amsmath}   % for math
\usepackage{amsfonts}  
\usepackage{subfigure} % for subfigures
\usepackage{hyperref}% for urls
\begin{document}
\title{Parity Nonconservation in Strong Interactions}
\author{Vernon~Barger}
%\email{barger@pheno.physics.wisc.edu}
\affiliation{Department of Physics, University of Wisconsin, Madison, WI 53706 USA}

\author {Wai-Yee Keung}
%\email{keung@fnal.gov}
\affiliation{Department of Physics, University of Illinois, Chicago, IL 60607-7059 USA}

\author{Chiu-Tien Yu}
%\email{cyu27@wisc.edu}
\affiliation{Department of Physics, University of Wisconsin, Madison, WI 53706 USA}\date{\today}

\begin{abstract}
For top-quarks produced via the subprocess $q\bar q\to t\bar t$, the longitudinal t-quark polarization ($P_{\parallel}$) vanishes in QCD.  $P_{\parallel}$ can be measured by the angular distribution of the lepton in $t$-quark semileptonic decay. New physics contributions that are parity nonconserving will be manifest by non-vanishing $P_{\parallel}$, which may be large.  We illustrate this with the $s$-channel exchange of a massive $X$-gluon with chiral quark couplings.
\end{abstract}
\maketitle

\renewcommand{\thefootnote}{\alph{footnote}}

%\section*{Introduction}
Quantum Chromodynamics (QCD), the gauge field theory describing the strong interactions of colored quarks and gluons in the Standard Model\cite{Gross:1973id}, has been extraordinarily successful in describing physics in both non-perturbative and perturbative regimes. Using the positivity of the Euclidean path integrand for Yang-Mills theory, Vafa and Witten proved that QCD does not spontaneously break parity or CP if $\bar\theta=0$\cite{Vafa:1984xg}. CP conservation in the strong interactions is necessitated by the extreme smallness of experimental upper bounds on the neutron electric dipole moment\footnotemark[1]{\footnotetext[1]{There are terms in the QCD Lagrangian that violate the charge-parity (CP) symmetry. Mechanisms have been proposed to solve this strong CP problem, of which the Peccei-Quinn mechanism is the most compelling \cite{Peccei:1977hh}.}. It has been suggested that heavy ion collisions may form metastable phases which allow for parity nonconservation\cite{Kharzeev:1998kz}.  An induced charge separation along the angular momentum vector of the collision would create an electric dipole moment of the hot gluon matter. There are ongoing searches by the STAR collaboration\cite{:2009txa} at the Relativistic Heavy Ion Collider (RHIC) to establish such an effect. Parity-violating effects can also be induced by topological solutions in QCD.

There have been many new physics models proposed to explain the large forward-backward asymmetry, $A_{FB}$, in top quark pair production seen at the Tevatron\cite{Aaltonen:2011kc}. For example, recent works have shown that an axial gluon\cite{Bai:2011ed}\cite{Frampton:2009rk}\cite{Tavares:2011zg}\cite{Haisch:2011up} can provide an explanation for the $A_{FB}$ measurement. For recent reviews of the many new physics models, see e.g. Ref.\cite{Bai:2011ed}\cite{Cao:2010zb}. However, $A_{FB}$ is a test of charge-conjugation (at tree-level) and not of parity conservation. Instead, one can look at the longitudinal polarization of the top-quark, which is a quantity solely determined by parity nonconservation that can be measured in collider experiments. A model that can lead to observable parity nonconservation is the $s$-channel exchange of a spin-1 $X$-gluon with both vector and axial-vector couplings to quarks\cite{Frampton:2009rk}\cite{Frampton:1987dn}, which we will use as an illustrative example in this Letter. The importance of the measurements of the longitudinal top-quark polarization has also be noted by other authors\cite{Kane:1991bg}\cite{Krohn:2011tw}. At all orders in perturbation theory, QCD leads to zero longitudinal polarization, and SM electroweak contributions should at most be at the few percent level.  Thus, the longitudinal top polarization is free of QCD theory ambiguities, unlike the case for the forward-backward asymmetry or the transverse component of the top polarization, both of which have QCD contributions.
%Recent works have shown that an axial gluon\cite{Bai:2011ed}\cite{Frampton:2009rk}\cite{Tavares:2011zg} can provide an explanation for the large forward-backward asymmetry, $A_{FB}$, seen at the Tevatron\cite{Aaltonen:2011kc}\cite{Abazov:2011rq}. However, the $A_{FB}$ is a test of charge-conjugation and not of parity-violation. Instead, one can look at the longitudinal polarization of the top-quark, which is a quantity solely determined by parity-violation and can be measured in collider experiments. There are new physics models that can lead to parity violating effects in the strong sector that can be observed at colliders, and searches for such effects are a potential way to detect new physics. For recent reviews of the many new physics models, see e.g. Ref.\cite{Cao:2010zb}\cite{Gresham:2011fx}\cite{Bai:2011ed}. An example of a model that can lead to observable parity violation is the $s$-channel exchange of a spin-1 $X$-gluon with both vector and axial-vector couplings to quarks. The importance of the measurements of the top-quark polarization has also be noted by other authors\cite{Kane:1991bg}\cite{Krohn:2011tw}. At all orders in perturbation theory, QCD leads to zero longitudinal polarization, and SM electroweak contributions should at most be at the few percent level.  Thus, the longitudinal top polarization is free of QCD theory ambiguities, unlike the case for the forward-backward asymmetry or the transverse component of the top polarization, both of which have QCD contributions.

%\section*{$X$-gluon Model}
{\bf{$X$-GLUON MODEL}}
Let $A_1$ and $A_2$ be non-abelian gauge fields associated with the
gauge group product, $SU_1(3) \times SU_2(3)$. The full symmetry is
broken by a bi-fundamental Higgs field $\Phi$ with a vev of the form $ \langle \Phi \rangle = V {\mathbf1} $.
The surviving gauge symmetry is the vectorial $SU_V(3)$. 
Since $T_2 |0\rangle = -T_1|0\rangle$, when the generators act upon the vev state, the massive $X$-gluon composition is
\begin{equation}X= (g_1 A_1 -g_2 A_2) / \sqrt{g_1^2+g_2^2}  \ ,\end{equation}
which has been normalized. The other orthogonal combination 
is the unbroken massless gluon field,
\begin{equation} G = (g_2 A_1 + g_1 A_2) / \sqrt{g_1^2+g_2^2}   \ . \end{equation}
\underline{$X$-gluon couplings to quarks}
The couplings to the generators are
\begin{eqnarray}g_1 A_1 T_1  + g_2 A_2 T_2
=\tfrac{1}{2} g_1 g_2/g_X G (T_1+T_2) \\
+ \tfrac{1}{2}g_X X (g_1^2 T_1- g_2^2 T_2)\nonumber.\end{eqnarray}
%%%
A further simplification gives
\begin{eqnarray}g_s G (T_1+T_2) 
     +\tfrac{1}{4}(g_1^2-g_2^2)/g_X X (T_1+T_2)\\ 
     + g_X X (T_1-T_2)\nonumber. \end{eqnarray}
where we define $g_X=\frac{1}{2} \sqrt{g_1^2+g_2^2}$ and $ g_s=\tfrac{1}{2}g_1g_2/g_X$. We set $T_1$ to act on $L$ chiral fields and $T_2$ on $R$ such that
$T_1+T_2$ acts only on the vectorial current, and 
$T_1-T_2$ on the axial-vectorial current: $\mathbf{T}_1+\mathbf{T}_2  \longrightarrow  \bar q\ \mathbf{T}\ \gamma^\mu q,~\mathbf{T}_1-\mathbf{T}_2  \longrightarrow - \bar q\ \mathbf{T}\ \gamma^\mu\gamma_5 q.$
The $X$-gluon interaction Lagrangian is
%%%
\begin{equation}   \mathbf{X}\cdot \bar q\ \mathbf{T\ }\gamma^\mu   (g_V^q+g_A^q\gamma_5) q \end{equation}
with $t\in q$ and $g_V^2=g_A^2-g_s^2$. This relationship of the couplings  is modified if one considers higher dimension operators\cite{Tavares:2011zg}
\begin{eqnarray}
{\cal L}& \supset&  \Lambda^{-2} \left[
  \lambda_Q^2 (\bar Q_L \Phi )   i\not\!\!D  (\phi^\dagger Q)
+\lambda_U^2 (\bar U_R \Phi^\dagger )   i\not\!\!D  (\phi U_R)\right.\nonumber \\
&+&\left. \lambda_D^2 (\bar D_R \Phi^\dagger )   i\not \!\!D  (\phi D_R)  \right]
\end{eqnarray}
The vev of the bi-fundamental Higgs $\phi$ allows the left-handed gauge field to act upon the right handed quark,
and vice versa such that $A_1  \hbox{ acts on } \mathbf{T_1} + y_{U|D}   \mathbf{T_2}$ and $A_2  \hbox{ acts on } \mathbf{T_2} + x         \mathbf{T_1}$,
%%%
where $x=\lambda_Q^2 V^2/\Lambda^2$ and $y_{U|D}=\lambda_{U|D}^2 V^2/\Lambda^2$. We also have
\begin{equation}\mathbf{T_1}\pm\mathbf{T_2} \Rightarrow (1\pm x)\mathbf{T_1}\pm(1\pm y)\mathbf{T_2}.\end{equation}
%%%%%
The kinetic derivative piece $i \! \not\!\!\partial$ is increased by $1+x$ for $Q$ and
$1+y_{U|D}$ for $U$ or $D$ respectively.
After renormalizing the kinetic pieces, we have 
\begin{eqnarray}
g^q_A&=&-\frac{g_X}{2}
      \left( \frac{1-x}{1+x}+ \frac{1-y_q}{1+y_q}  \right)  \nonumber,\\ 
%%%
g^q_V&=& \frac{g_1^2-g_2^2}{ 4 g_X} +\frac{g_X}{2}
  \left(  \frac{1-x}{ 1+x}-\frac{1-y_q}{ 1+y_q}  \right).
  \end{eqnarray}
The restrictions on the couplings noted above disappear when higher dimension operators are included.

\underline{$q\bar q\to t\bar t$}
The helicity amplitudes for the subprocess $q\bar q\to t\bar t$, shown in Fig.\ref{feynmandiagram}, are given in Table \ref{helicityamplitudes}, where $\theta$ is the CM scattering angle, $\beta^2=1-4m_t^2/\hat s$, and 
\begin{eqnarray} G_{I,V}
&=& \frac{g_s^2}{ \hat s} + \frac{ g_I^q g_V^t }{ \hat s-m_{X}^2+im_{X}\Gamma_{X}},\nonumber\\
G_{I,A}
&=&  \frac{ g_I^q g_A^t }{\hat s-m_{X}^2+im_{X}\Gamma_{X}}\end{eqnarray}
%%%
The $G_{I,V(A)} $ are functions of $\hat s$, the square of the subprocess center-of-mass energy, and carry two subscripts; 
The first refers to 
initial quark chiralities, $I=L$ or $R$ where we define $g_L=\tfrac{1}{2}(g_V-g_A)$, $g_R=\tfrac{1}{2}(g_V+g_A)$. The massless condition on 
initial quarks simplifies the calculation with
couplings in this basis. The second subscript refers to the
vectorial or axial-vectorial nature of the top quark couplings, which
is more efficient in dealing with massive states.
\begin{figure*}[htbp]
\begin{center}
\includegraphics[scale=0.6]{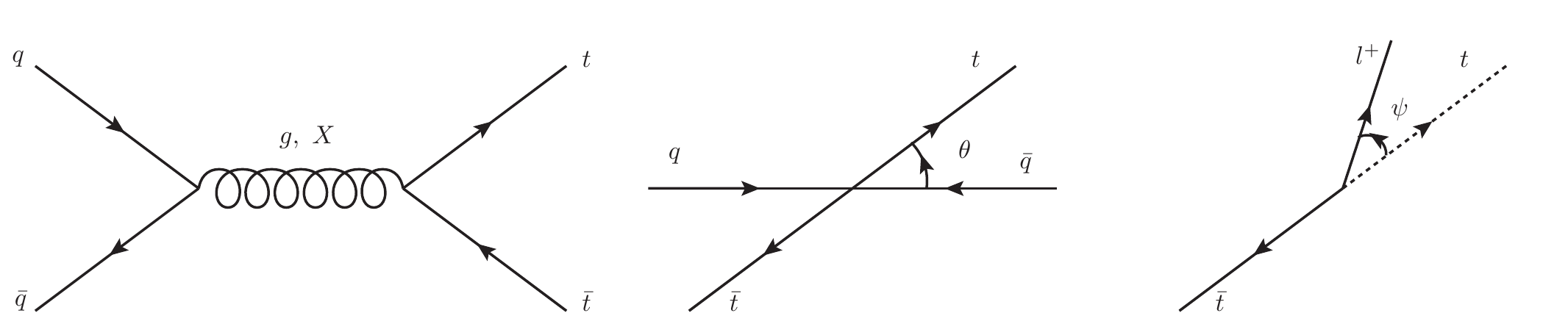}
\caption{Feynman diagram of the $s$-channel exchanges of the gluon and $X$-gluon in the $q\bar q\to t\bar t$ subprocess and the definitions of $\theta$ and $\psi$. The dotted line for $t$ denotes a boost into the $t$ rest frame.}
\label{feynmandiagram}
\end{center}
\end{figure*}

\begin{table*}[htdp]
\caption{Helicity amplitudes for $q\bar q\to t\bar t$}
\begin{center}
\renewcommand{\arraystretch}{1.3}
\begin{tabular}{|c||c|c|}
\hline
&\multicolumn{2}{c|}{Initial State Polarizations}\\
Final State Polarizations&\multicolumn{1}{c}{$-+$}&\multicolumn{1}{c|}{$+-$}\\\hline\hline
$--$&$G_{L,V} 2\sqrt{\hat s} m_t \sin\theta $&$ G_{R,V} 2\sqrt{\hat s} m_t \sin\theta$\\ \hline
$++$&$-G_{L,V} 2\sqrt{\hat s} m_t \sin\theta $&$ -G_{R,V} 2\sqrt{\hat s} m_t \sin\theta$  \\\hline
$-+$&$-(G_{L,V}-\beta G_{L,A}) \hat s (1+\cos\theta) $&$ (G_{R,V}-\beta G_{R,A})\hat s (1-\cos\theta) $ \\\hline
$+-$&$(G_{L,V}+\beta G_{L,A})\hat s (1-\cos\theta)  $&$ -(G_{R,V}+\beta G_{R,A})\hat s (1+\cos\theta) $ \\\hline
\end{tabular}
\end{center}
\label{helicityamplitudes}
\end{table*}%

%In the heavy gluon model, the couplings are defined by
%\begin{equation}{\cal L} \supset \bar q \gamma_\mu(g_V^q+g_A^q\gamma_5) q X^\mu\end{equation}
%where we assume that $t\in q.$ 
We define $\tilde\sigma(\theta)=\sum|{\cal{M}}|^2(q\bar q\to t\bar t)$, which is the subprocess differential cross-section modulo an overall factor\footnotemark[2]{\footnotetext[2]{$\frac{d\hat\sigma}{d\cos\theta}=\frac{\beta}{576\pi\hat s}\tilde\sigma(\theta)$.}}. Then, we have
%\begin{eqnarray}
% \sum |{\cal M}|^2
%&=&\ s^2 (|G_{L,V}|^2 + |G_{R,V}|^2)2(1+\cos^2\theta+4m_t^2\sin^2\theta/s) 
%                                                          \nonumber\\
% &+& 8 s^2\hbox{ Re}(G^*_{L,V}G_{L,A}- G^*_{R,V}G_{R,A})
%                            \beta\cos\theta         \nonumber\\
% &+& s^2 (|G_{L,A}|^2 + |G_{R,A}|^2)2(1-4m_t^2/s) (1+\cos^2\theta) 
% \label{M2}
%\end{eqnarray}
\begin{eqnarray}
 \label{M2}
\tilde\sigma(\theta)
&=&[A^+(-\beta)+A^+(\beta)](1+\cos^2\theta)\nonumber\\
&-&2[A^-(\beta)-A^-(-\beta)]\cos\theta\nonumber\\
&+&2(1-\beta^2)A^+(0)\sin^2\theta
\end{eqnarray}
where 
\begin{equation}
A^{\pm}(\beta)=\hat s^2\left(|G_{L,V}+\beta G_{L,A}|^2\pm|G_{R,V}+\beta G_{R,A}|^2\right).
\end{equation}
 Note that the second line of Eq.\ref{M2} gives rise to the forward-backward asymmetry. Since the $s$-channel gluon and $X$-gluon amplitudes have identical color structure, the polarization and asymmetry predictions are independent of the parton distribution functions.

For the 7 TeV run of the LHC (LHC7), the analysis is complicated by subprocesses that involve gluons as partons. By a selective choice of the rapidity region that emphasizes the $q\bar q\to t\bar t$ subprocess, it may be possible to probe parity and $C$ nonconservation at the LHC\cite{Krohn:2011tw}, as well as at the Tevatron. 
%However, by considering a high rapidity band, 
%$2<|y_{t\bar t}|$, which corresponds to an angular range of $\theta<15^\circ \mathrm{~or~}165^\circ<\theta$ where the $q\bar q\to t\bar t$ subprocess contribution is dominant, it may be possible to probe parity and $C$ nonconservation at the LHC, as well as at the Tevatron. 

%\section*{Top Quark Polarization}
{\bf{LONGITUDINAL POLARIZATION OF TOP}}
For our purposes, we will consider the leading-order production of top quarks. To all orders of QCD, the top quarks produced are unpolarized. However, in $X$-gluon models, the chiral structure gives rise to partially polarized tops. The longitudinal polarization of the top is described by
%%

%\begin{equation}P_\parallel =
%\frac{\sum_{h_q, h_{\bar q}, h_{\bar t}}
%|(h_q, h_{\bar q}, + , h_{\bar t})|^2-|(h_q, h_{\bar q}, - , h_{\bar t})|^2
%}{
%\sum_{h_q, h_{\bar q}, h_{\bar t}}
%|(h_q, h_{\bar q}, + , h_{\bar t})|^2+|(h_q, h_{\bar q}, - , h_{\bar t})|^2
%}
%\end{equation}
\begin{equation}
P_\parallel =
\frac{\sum\left[|(h_q, h_{\bar q}, + , h_{\bar t})|^2-|(h_q, h_{\bar q}, - , h_{\bar t})|^2\right]
}{
\sum\left[|(h_q, h_{\bar q}, + , h_{\bar t})|^2+|(h_q, h_{\bar q}, - , h_{\bar t})|^2\right]
}=\frac{\tilde\sigma_{\parallel}(\theta)}{\tilde\sigma(\theta)}
\end{equation}
where the sum is over helicities and $\tilde\sigma(\theta)$ is given by Eq. \ref{M2}. The numerator can be simplified as
\begin{eqnarray}
\label{eq:pp}
\tilde\sigma_{\parallel}(\theta)&=&[A^+(\beta)-A^+(-\beta)](1+\cos^2\theta)\nonumber\\
&-&2[A^-(-\beta)+A^-(\beta)]\cos\theta
\end{eqnarray}
%In the case of a $V-A~(g_V=-g_A)$ coupling, Eqs. \ref{M2} and \ref{eq:pp} simplify to
%\begin{eqnarray}
%\hat\sigma_{tot}&=&2|G_{L,V}|^2\left[(1+\beta^2)(1+\cos^2\theta)+4\beta\cos\theta\right.\nonumber\\
%&-&\left.(1-\beta^2)\sin^2\theta\right]\\
%\hat\sigma_{P_\parallel}&=&2|G_{L,V}|^2\left[2\beta(1+\cos^2\theta)-2\cos\theta\right.\nonumber\\
%&-&\left.(1-\beta^2)\sin^2\theta\right].
%\end{eqnarray}

%\begin{eqnarray}
%\hat\sigma_{P_{\parallel}}&=4s^2\cos\theta(|G_{R,V}|^2+\beta^2|G_{R,A}|^2
%              -|G_{L,V}|^2-\beta^2|G_{L,A}|^2)\nonumber\\
%%%%
%&- 4s^2 \beta\hbox{ Re }(G_{L,V}^*G_{L,A}+ G_{R,V}^*G_{R,A})(1+\cos^2\theta).
%\end{eqnarray}
%%
The antilepton $\ell^+$ from the top decay has an
angular distribution given by  $(1+P_\parallel \cos\psi)$, where $\psi$ is defined in Eq.\ref{pp}. 
We obtain a similar expression for $\bar t$ by exchanging $h_t\leftrightarrow h_{\bar t}$.
%\begin{equation} \bar P_\parallel =
%\frac{\sum_{h_q, h_{\bar q}, h_{ t}}
%|(h_q, h_{\bar q}, h_{t}, +)|^2-|(h_q, h_{\bar q} , h_{t},-)|^2}{\sum_{h_q, h_{\bar q}, h_{\bar t}}
%|(h_q, h_{\bar q}, h_{t}, +)|^2+|(h_q, h_{\bar q} , h_{t},-)|^2}
%\end{equation}
%%%
The corresponding angular distribution of $\ell^-$ is 
$1-\bar P_\parallel \cos\bar\psi$.
From CP-nonconservation, $P_\parallel=-\bar
P_\parallel$, so that the angular distributions of $\ell^\pm$ 
are symmetric under $CP$.
 
The angle $\psi$ is defined as the angle between the $\ell^+$ and the negative $\bar t$ momentum in a boosted frame in which the $t$ is at rest.
The Lorentz boost to the $t\bar t$ CM  frame gives
\begin{equation}E_{t\bar t}(e^+)=E_{t}(e^+) (1+\beta\cos\psi)\left(\tfrac{1}{2}M_{t\bar t}/m_t\right)\end{equation}
so that
\begin{equation} (2m_t/M_{t\bar t})p_{\ell^+}\cdot (p_t+p_{\bar t}) /M_{t\bar t}
=p_{\ell^+}\cdot p_t (1+\beta\cos\psi)/m_t \end{equation}
\begin{equation}\cos\psi=\left[
\frac{ 2m_t^2}{M_{t\bar t}^2}
\left(\frac{p_{\ell^+}\cdot p_{\bar t}}{ p_{\ell^+}\cdot p_t }+1\right)-1 \right]
\left /\sqrt{1-\frac{4m_t^2}{M_{t\bar t}^2} }\right.\label{pp}\end{equation}
The expression above makes use of covariant 4-dot-products and can be evaluated in any frame.  Here, we assume all momenta can be
reconstructed in the experiment.
$\bar \psi$ is obtained from Eq.\ref{pp} with substitutions $\psi\to\bar\psi$,
$\ell^+\to \ell^-$, and $t\leftrightarrow \bar t$. 
%Similarly, the angle $\bar\psi$ is defined at the $\bar t$ rest frame 
%as the angle between $\ell^-$ and the direction from the recoiling $t$ 
%to the rest $\bar t$.
%%%
%\begin{equation} \cos\bar\psi=\left[
%\frac{ m_t}{M_{t\bar t}}
%\left(\frac{p_{\ell^-}\cdot p_{ t}}{p_{\ell^-}\cdot p_{\bar t} }
%+1\right) -1\right]
%\left /\sqrt{1-\frac{4m_t^2}{M_{t\bar t}^2}} \right.\end{equation}

%\section*{Phenomenology}
{\bf{PHENOMENOLOGY}}
%\underline{Tevatron}
For our illustrations, we adopt the parameters $|g_A|=g_s/3, M_X=420$ GeV, and $\Gamma_X=42$ GeV of the $A_{FB}$ model of Ref.\cite{Tavares:2011zg}, but we allow for the possibility of a vector coupling as well, which leads to parity nonconservation.  For simplicity, we consider maximal parity nonconservation scenarios, which we denote as $V\pm A.$ The new physics contribution to the $M_{t\bar t}$ distribution is found to be similar in the $V\pm A$ cases to that for $A$-only in Ref.\cite{Tavares:2011zg}, as shown in Fig.\ref{mtt}. We do not take into account smearing due to the experimental $M_{t\bar t}$ resolution. We show the dependence of the $A_{FB}$ on top-pair invariant mass $M_{t\bar t}$ in Fig. \ref{fig:afbmtt}.
% and the invariant mass distribution in Fig.\ref{mtt}. We consider two combinations of couplings: (i) $g_X=g_s/3$, we use $M_{X}=420$ GeV and $\Gamma_{X}=42$ GeV\cite{Tavares:2011zg} and (ii) $g_X=g_s$, $M_{X}=1000$ GeV and $\Gamma_{X}=100$ GeV\cite{Bai:2011ed}. We see that at high invariant mass, the axial-only coupling leads to a larger positive asymmetry, whereas the $V-A$ and $V+A$ couplings lead to a larger asymmetry at lower invariant mass. In case (i), the asymmetry created by the $V\pm A$ couplings is always positive, whereas the axial-coupling leads to a negative asymmetry at low invariant mass. For case (ii), both the $V\pm A$ and axial-only couplings give a negative asymmetry at low invariant mass and positive asymmetry at high invariant mass. 

%\begin{figure*}[htbp]
%\begin{center}
%\subfigure[~$g_X=g_s/3, M_{X}=420$ GeV, $\Gamma_{X}=42$ GeV.]{
%\includegraphics[scale=0.8]{afb_mtt_gsga3_mgx420}
%}
%\subfigure[~$g_X=g_s, M_{X}=1000$ GeV, $\Gamma_{X}=100$ GeV.]{
%\includegraphics[scale=0.8]{afb_mtt_gsga_mgx1000}}
%\caption{$A_{FB}$ vs. $M_{t\bar t}$ at $M_{X}=420, 1000$ GeV, $\Gamma_{X}=42, 100$ GeV for $V\pm A$, Axial-only, and Vector-only couplings.}
%\label{fig:afbmtt}
%\end{center}
%\end{figure*}
\begin{figure}[htbp]
\begin{center}
\includegraphics[scale=0.86]{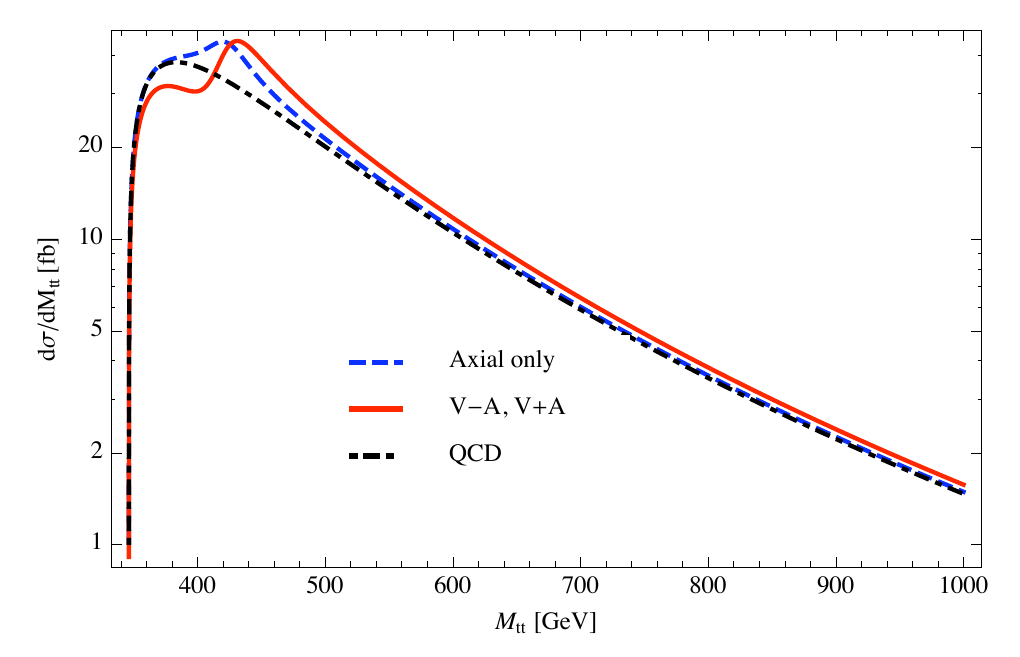}
%\includegraphics[scale=0.85]{dsdmtt_gsga3_mgx420_pdf_norm_v2}
%\caption{The new physics (NP) contributions to $d\sigma/dM_{t\bar t}$ vs. $M_{t\bar t}$ in the $X$-gluon model at the Tevatron, normalized to the SM.}
\caption{The new physics (NP) contributions to $d\sigma/dM_{t\bar t}$ vs. $M_{t\bar t}$ in the $X$-gluon model at the Tevatron.}
\label{mtt}
\end{center}
\end{figure}

\begin{figure*}[htbp]
\begin{center}
\subfigure[$A_{FB}$ vs. $M_{t\bar t}$ for $V\pm A$~(solid), Axial-only~(dashed), and Vector-only~(dotted) couplings.]{\label{fig:afbmtt}\includegraphics[scale=0.87]{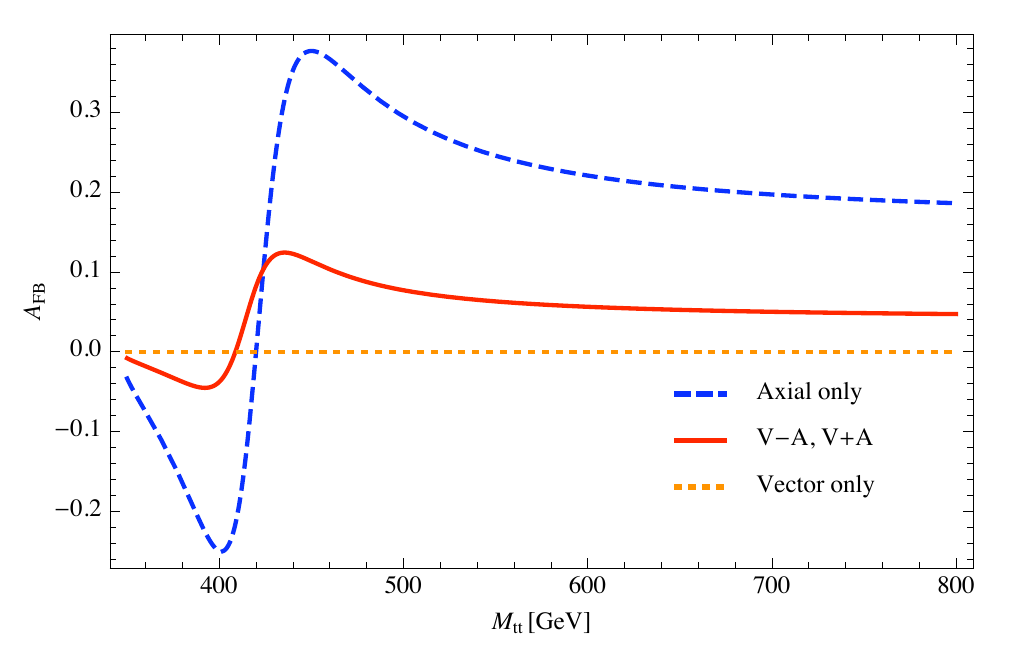}}
\subfigure[$P_{||}$ vs. $M_{t\bar t}$ for $\theta=0~(\mathrm{dashed})\mathrm{~and~}\pi~({\mathrm{solid}})$. $P_\parallel$ for $V+A$ is opposite in sign to $P_\parallel$ for $V-A$.]{\includegraphics[scale=0.87]{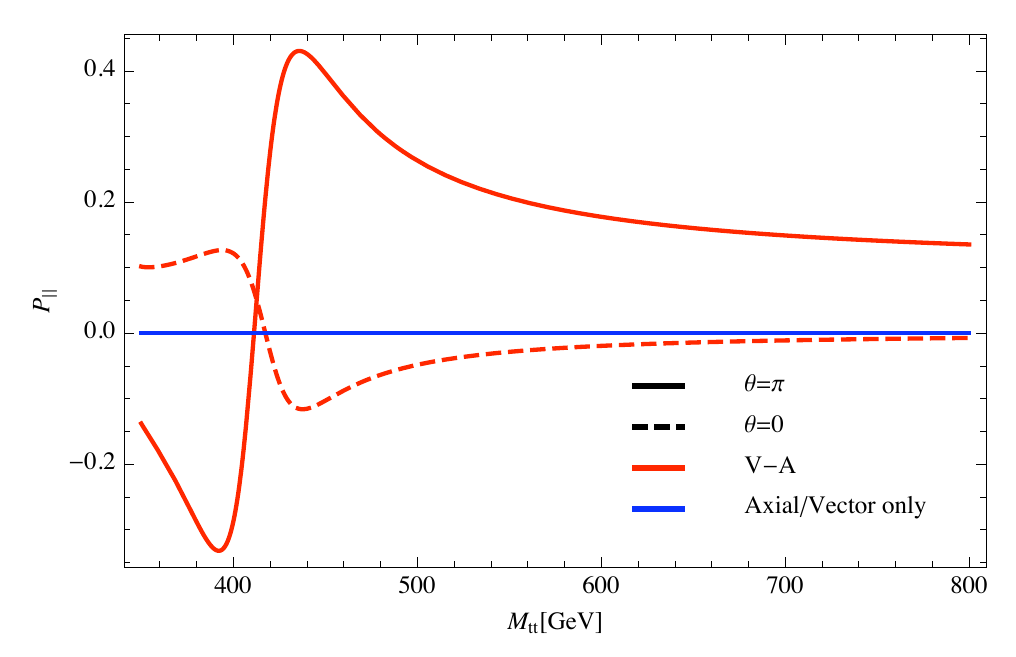}
\label{fig:pp_mtt}
}
\caption{$A_{FB}$  and $P_{||}$ vs. $M_{t\bar t}$ for $M_{X}=420$ GeV, $\Gamma_{X}=42$ GeV, and $|g_A|=g_s/3$.}
\label{fig:afbpp}
\end{center}
\end{figure*}

%\underline{LHC}
%\begin{figure*}[htbp]
%\begin{center}
%\subfigure[~$g_X=g_s/3, M_{X}=420$ GeV, $\Gamma_{X}=42$ GeV.]{\includegraphics[scale=0.72]{pp_theta_gsga_mgx420_reverse}}
%\subfigure[~$g_X=g_s, M_{X}=1000$ GeV, $\Gamma_{X}=100$ GeV.]{\includegraphics[scale=0.81]{pp_theta_gsga_mgx1000_reverse_nolegend}}
%\caption{$P_{||}$ vs. $\theta$ for $M_{X}=420, 1000$ GeV, $\Gamma_{X}=42, 100$ GeV and $\sqrt{\hat s}=500, 1000$ GeV.}
%\label{fig:pp_theta}
%\end{center}
%\end{figure*}
%We show the dependence of $P_{\parallel}$ on the CM scattering angle $\theta$ in Fig. \ref{fig:pp_theta}. The polarization is greater for smaller $\sqrt{\hat s}$. Increasing the mass of the $X$-gluon also leads to larger polarization.  As expected, there is no polarization for a purely axial or vector coupling. 
We show the dependence of $P_{\parallel}$ on $M_{t\bar t}$ in Fig. \ref{fig:pp_mtt}. The polarization goes to zero near $M_{t\bar t}=M_X$. $P_\parallel$ is largest for $\theta=\pi$. Zero polarization is predicted for a purely axial or vector coupling.

\underline{Tri-gauge Boson Couplings~} The trilinear couplings in the $X$-gluon model are
\begin{equation}f_{ijk} \
\left\langle g_s G^i G^j G^k     
          +  g_s G^i X^j X^k
     +  2g_X X^i X^j X^k
   \right\rangle \end{equation}
   where the bra-ket denotes
%%%
\begin{eqnarray} &\langle \phi^i \phi^j \phi^k \rangle
=  \phi^i_\mu  \phi^j_\nu  \partial^\mu  {\phi^k\ }^\nu 
+  \phi^j_\mu  \phi^k_\nu  \partial^\mu  {\phi^i\ }^\nu 
+  \phi^k_\mu  \phi^i_\nu  \partial^\mu  {\phi^j\ }^\nu\nonumber\\
&- \phi^i_\mu  \phi^k_\nu  \partial^\mu  {\phi^j\ }^\nu
-  \phi^j_\mu  \phi^i_\nu  \partial^\mu  {\phi^k\ }^\nu
-  \phi^k_\mu  \phi^j_\nu  \partial^\mu  {\phi^i\ }^\nu\end{eqnarray}
which gives antisymmetric dependence on momenta. There is also a 4-particle coupling.
Thus, the $X$-gluon can be strongly pair-produced by gluon fusion via an $s$-channel gluon. Searches for the $X$-gluon are dependent on its decay modes, which are multijets in the scenario of Ref.\cite{Tavares:2011zg}. It is necessary to measure the tri-gauge couplings of the $X$-gluon to prove its $SU(3)$ color property.

%\section*{Conclusions}
{\bf{Conclusions}}
For top-quarks produced via the subprocess $q\bar q\to t\bar t$, the longitudinal t-quark polarization ($P_{\parallel}$) vanishes in QCD. As a new physics illustration, we have shown that the $s$-channel exchange of a massive $X$-gluon with chiral quark couplings gives rise to a substantial $P_{\parallel}$. Our study emphasizes the low-energy phenomenology and its parity nonconservation. Additional fermions are needed for UV completion and to cancel anomalies. The longitudinal polarization is a measurement of ${\mathbf{\sigma}}_t\cdot
(\mathbf{p}_{t}-\mathbf{p}_{\bar t}) /|\mathbf{p}_{t}-\mathbf{p}_{\bar t}|$ in the $t$
rest frame. Being odd in spatial parity, it is expected to be zero in
all orders of perturbative QCD. Thus measurements of $P_\parallel$ in
$t\bar t$ events arising from the $q\bar q\to t\bar t$ subprocess at
the Tevatron and LHC could prove to be of fundamental importance in
finding parity-violation in strong interactions beyond QCD.
%$P_{\parallel}$ can be measured by the angular distribution of the lepton in $t$-quark semileptonic decay. Thus, measurements of $P_\parallel$ in $t\bar t$ events arising from the $q\bar q\to t\bar t$ subprocess at the Tevatron and LHC could prove to be of fundamental importance in finding parity nonconservation in strong interactions beyond QCD.

{\bf{Acknowledgements}}
%\section*{Acknowledgements} 
The authors wish to thank Dan Amidei and William Murdock for their helpful communication. This work was supported in part by the U.S. Department of Energy under grants Nos. DE-FG02-95ER40896 and DE-FG02-84ER40173, and by the National Science Foundation under Grant No. NSF PHY05-51164.  C.-T.Y is supported by a NSF Graduate Research Fellowship. VB and C.-T.Y thank the KITP-Santa Barbara for its hospitality.  

\begin{thebibliography}{xx}
%
%\bibliography{bib}
%\bibliographystyle{unsrt}
%%
%%
%\refpta

%\cite{Gross:1973id}
\bibitem{Gross:1973id}
  D.~J.~Gross, F.~Wilczek,
  %``Ultraviolet Behavior of Nonabelian Gauge Theories,''
  Phys.\ Rev.\ Lett.\  {\bf 30}, 1343-1346 (1973).
  %\cite{Politzer:1973fx}
%\bibitem{Politzer:1973fx}
  H.~D.~Politzer,
  %``Reliable Perturbative Results for Strong Interactions?,''
  Phys.\ Rev.\ Lett.\  {\bf 30}, 1346-1349 (1973).
  
%\cite{Vafa:1984xg}
\bibitem{Vafa:1984xg}
  C.~Vafa, E.~Witten,
  %``Parity Conservation in QCD,''
  Phys.\ Rev.\ Lett.\  {\bf 53}, 535 (1984).
  
  %\cite{Peccei:1977hh}
\bibitem{Peccei:1977hh}
  R.~D.~Peccei, H.~R.~Quinn,
  %``CP Conservation in the Presence of Instantons,''
  Phys.\ Rev.\ Lett.\  {\bf 38}, 1440-1443 (1977).
 
% %\cite{Baker:2006ts}
%\bibitem{Baker:2006ts}
%  C.~A.~Baker, D.~D.~Doyle, P.~Geltenbort, K.~Green, M.~G.~D.~van der Grinten, P.~G.~Harris, P.~Iaydjiev, S.~N.~Ivanov {\it et al.},
%  %``An Improved experimental limit on the electric dipole moment of the neutron,''
%  Phys.\ Rev.\ Lett.\  {\bf 97}, 131801 (2006).
%  [hep-ex/0602020].
  
%  %\cite{Lee:1974ma}
%\bibitem{Lee:1974ma}
%  T.~D.~Lee, G.~C.~Wick,
%  %``Vacuum Stability and Vacuum Excitation in a Spin 0 Field Theory,''
%  Phys.\ Rev.\  {\bf D9}, 2291 (1974).
  
%\cite{Kharzeev:1998kz}
\bibitem{Kharzeev:1998kz}
  D.~Kharzeev, R.~D.~Pisarski, M.~H.~G.~Tytgat,
  %``Possibility of spontaneous parity violation in hot QCD,''
  Phys.\ Rev.\ Lett.\  {\bf 81}, 512-515 (1998).
  [hep-ph/9804221].
 
 %\cite{:2009txa}
\bibitem{:2009txa}
  B.~I.~Abelev {\it et al.} [ STAR Collaboration ],
  %``Observation of charge-dependent azimuthal correlations and possible local strong parity violation in heavy ion collisions,''
  Phys.\ Rev.\  {\bf C81}, 054908 (2010).
  [arXiv:0909.1717 [nucl-ex]]. 
%\cite{:2009uh}
%\bibitem{:2009uh}
%  B.~I.~Abelev {\it et al.} [ STAR Collaboration ],
%  %``Azimuthal Charged-Particle Correlations and Possible Local Strong Parity Violation,''
%  Phys.\ Rev.\ Lett.\  {\bf 103}, 251601 (2009).
%  [arXiv:0909.1739 [nucl-ex]].

  %\cite{Aaltonen:2011kc}
\bibitem{Aaltonen:2011kc}
  T.~Aaltonen {\it et al.} [ CDF Collaboration ],
  %``Evidence for a Mass Dependent Forward-Backward Asymmetry in Top Quark Pair Production,''
  Phys.\ Rev.\  {\bf D83}, 112003 (2011).
  [arXiv:1101.0034 [hep-ex]].
  %\cite{Abazov:2011rq}
\bibitem{Abazov:2011rq}
  V.~M.~Abazov {\it et al.} [ D0 Collaboration ],
  %``Forward-backward asymmetry in top quark-antiquark production,''
  [arXiv:1107.4995 [hep-ex]].  
     %\cite{Bai:2011ed}
\bibitem{Bai:2011ed}
  Y.~Bai, J.~L.~Hewett, J.~Kaplan, T.~G.~Rizzo,
  %``LHC Predictions from a Tevatron Anomaly in the Top Quark Forward-Backward Asymmetry,''
  JHEP {\bf 1103}, 003 (2011).
  [arXiv:1101.5203 [hep-ph]].
 %\cite{Frampton:2009rk}
\bibitem{Frampton:2009rk}
  P.~H.~Frampton, J.~Shu, K.~Wang,
  %``Axigluon as Possible Explanation for p anti-p ---> t anti-t Forward-Backward Asymmetry,''
  Phys.\ Lett.\  {\bf B683}, 294-297 (2010).
  [arXiv:0911.2955 [hep-ph]].

    %\cite{Tavares:2011zg}
\bibitem{Tavares:2011zg}
  G.~M.~Tavares, M.~Schmaltz,
  %``Explaining the t-tbar asymmetry with a light axigluon,''
  [arXiv:1107.0978 [hep-ph]].
  %\cite{Haisch:2011up}
\bibitem{Haisch:2011up}
  U.~Haisch, S.~Westhoff,
  %``Massive Color-Octet Bosons: Bounds on Effects in Top-Quark Pair Production,''
  [arXiv:1106.0529 [hep-ph]].
  %\cite{Cao:2010zb}
\bibitem{Cao:2010zb}
  Q.~-H.~Cao, D.~McKeen, J.~L.~Rosner, G.~Shaughnessy, C.~E.~M.~Wagner,
  %``Forward-Backward Asymmetry of Top Quark Pair Production,''
  Phys.\ Rev.\  {\bf D81}, 114004 (2010).
  [arXiv:1003.3461 [hep-ph]].
 %\cite{Gresham:2011fx}
%\bibitem{Gresham:2011fx}
  M.~I.~Gresham, I.~-W.~Kim, K.~M.~Zurek,
  %``Tevatron Top $A_{FB}$ Versus LHC Top Physics,''
  [arXiv:1107.4364 [hep-ph]]. %\cite{AguilarSaavedra:2011ug}
%\bibitem{AguilarSaavedra:2011ug}
  J.~A.~Aguilar-Saavedra, M.~Perez-Victoria,
  %``Simple models for the top asymmetry: Constraints and predictions,''
  [arXiv:1107.0841 [hep-ph]].
  %\cite{Frampton:1987dn}
\bibitem{Frampton:1987dn}
  P.~H.~Frampton, S.~L.~Glashow,
  %``Chiral Color: An Alternative to the Standard Model,''
  Phys.\ Lett.\  {\bf B190}, 157 (1987).
%%\cite{Kane:1991bg}
\bibitem{Kane:1991bg}
  G.~L.~Kane, G.~A.~Ladinsky, C.~P.~Yuan,
  %``Using the Top Quark for Testing Standard Model Polarization and CP Predictions,''
  Phys.\ Rev.\  {\bf D45}, 124-141 (1992).  
    %\cite{Jung:2010yn}
%\bibitem{Jung:2010yn}
  D.~-W.~Jung, P.~Ko, J.~S.~Lee,
  %``Longitudinal top polarization as a probe of a possible origin of forward-backward asymmetry of the top quark at the Tevatron,''
  Phys.\ Lett.\  {\bf B701}, 248-254 (2011).
[arXiv:1011.5976 [hep-ph]].
     %\cite{Krohn:2011tw}
\bibitem{Krohn:2011tw}
  D.~Krohn, T.~Liu, J.~Shelton, L.~-T.~Wang,
  %``A Polarized View of the Top Asymmetry,''
  [arXiv:1105.3743 [hep-ph]]. 

%%
%\cite{Alvarez:2011hi}
%\bibitem{Alvarez:2011hi}
%  E.~Alvarez, L.~Da Rold, J.~I.~S.~Vietto, A.~Szynkman,
%%``Phenomenology of a light gluon resonance in top-physics at Tevatron and LHC,''
%%  
%  [arXiv:1107.1473 [hep-ph]].
%%%%
%%%\cite{Grzadkowski:1993rd}
%%\bibitem{Grzadkowski:1993rd}
%%  B.~Grzadkowski, W.~-Y.~Keung,
%%  %``SUSY induced CP violation in t decays at e- e+ colliders,''
%%  Phys.\ Lett.\  {\bf B316}, 137-147 (1993).
%%  [hep-ph/9306322].
%%\cite{Kamenik:2011wt}
%\bibitem{Kamenik:2011wt}
%  J.~F.~Kamenik, J.~Shu, J.~Zupan,
%  %``Review of new physics effects in t-tbar production,''
%  [arXiv:1107.5257 [hep-ph]].

\end{thebibliography}
\end{document}